\documentclass[twocolumn,showpacs,preprintnumbers,amsmath,amssymb]{revtex4}

\usepackage{graphicx}
\usepackage{dcolumn}
\usepackage{bm}

\def\pb{$\overline{p}\;$}
\def\pbs{$\overline{p}$s }

\def\nu{$\overline{N}\;$}

\begin{document}

\preprint{APS/123-QED}

\title{Parameterization of the antiproton inclusive production cross section on nuclei}
\author{R.P. Duperray, C.-Y. Huang\footnote{Present address: Max Planck Institute f\"ur 
Kernphysik, D-69117 Heidelberg.}, 
K.V. Protasov, and M. Bu\'enerd \footnote{Corresponding author: buenerd@lpsc.in2p3.fr }}
\affiliation{Laboratoire de Physique Subatomique et de Cosmologie \\
  53, avenue des Martyrs - 38026 Grenoble cedex, France}%
\date{\today}

\begin{abstract}
A new parameterization of the $\overline{p}$ inclusive production cross
section in proton-proton and proton-nucleus collisions is proposed. A sample of consistent
$pA\rightarrow \overline{p}X$ experimental data sets measured on $1\leq A \leq208$ nuclei,
from 12~GeV up to 400~GeV incident energy, have been used to constrain the parameters.
A broader energy domain is covered for the $pp\rightarrow \overline{p}X$ reaction and
with a simplified functional form used in the fits. The agreement obtained with the data is
good. The results are discussed.
\end{abstract}

\maketitle
%
%
\section{Introduction}
An accurate description of the inclusive antiproton production cross section in proton-nucleus 
collisions, necessarily relies on the empirical approach to the experimental data since
theoretical calculations can provide at best approximate values in the current stage of the 
theory. The aim of the present work was to develop a handy analytical parameterization for the 
description of the inclusive \pb production cross section in proton-proton ($pp$) and 
proton-nucleus ($pA$) collisions on the basis of the existing body of data, updating the former
works on the subject.

The motivations of the work have their origin in the needs of Cosmic Ray (CR) physics where a 
good knowledge and a good description of the \pb inclusive production cross section is a key 
requirement for a detailed understanding of the production and propagation of secondary galactic 
and atmospheric antiprotons.
The $\overline{p}$ component of the CR flux is an important window for cosmology. The main
contribution to this flux originates from the interaction of the high energy CR flux with
the interstellar matter (ISM) in the galaxy. These \pbs are called secondary galactic. In
addition to the secondary products, a primary component could exist,
undergone for example, by the dark matter neutralino annihilation \cite{JU94} or by the
evaporation products of primordial black holes \cite{BA00}, both being of major
physical and astrophysical interest. Such signatures could be obtained only if the basic
processes of galactic and atmospheric $\overline{p}$ production cross section in $pp$ and
$pA$ collisions are known with a good enough accuracy over
a momentum range extending from around threshold up to a few hundreds of GeV where the
CR flux becomes vanishingly small.

The approach used here closely follows the forms used by Kalinovskii, Mokhov and Nikitin
\cite{KMN} -- referred to as KMN in the following -- for the description of the
$pA\rightarrow \bar{p}X$ cross section. The functional form used in this reference has been
modified in order to reproduce a much larger sample of data sets over a much larger
dynamical range and for a larger range of nuclear mass, than in the original work. This
work extends a previous effort covering a more limited domain of incident momentum and of
nuclear mass \cite{HU02,HU03}.
%
%
\section{Inclusive cross sections in hadron collisions}\label{FEAT}
In high energy hadron collisions the final state is often complex, many particles being
produced in the collision. The inclusive single particle production cross section is a 
quantity of interest in many physics studies, for a reaction $ab \rightarrow cX$, where
$c$ is the particle of interest and where $X$ represents all the other particles produced 
in any quantum final state allowed in the collision. The invariant inclusive single particle 
distribution is defined by:
\begin{eqnarray}
f\left( ab\rightarrow cX\right) =E_{c}\frac{d^{3}\sigma }{dp_{c}^{3}}=\frac{
E_{c}}{\pi }\frac{d^{2}\sigma }{dp_{\parallel }dp_{\perp}^{2}}=\frac{d^{2}\sigma }{\pi
dyd\left( p_{\perp}^{2}\right)},
\label{eq1}
\end{eqnarray}
where $d^{3}\sigma /dp_{c}^{3}$ is the triple differential cross section for detecting
particle $c$ within the phase-space volume element $d^{3}p_{c}$. $E_{c}$ is the total
energy of $c$, while $p_{\parallel}$ and $p_{\perp}$ are the longitudinal and transverse
components of ${\mathbf{p}}_{c}$, respectively. The rapidity variable
$y=0.5\ln ((E+p_{\parallel})/(E-p_{\parallel}))$ is often used to describe the
$p_{\parallel}$ dependence of the cross section because of its interesting properties in
Lorentz transformations \cite{BYK}. To obtain two last expression in (\ref{eq1}),
an azimuthal symmetry of the differentiel cross section was used.
It is also convenient to introduce the following
dimensionless variables:
\begin{eqnarray}
x_{f}=\frac{p_{\parallel }^{\ast }}{p_{\parallel \max }^{\ast }} \mbox{ and }
x_{R}=\frac{E^{\ast }}{E_{\max }^{\ast }},
\label{eq2}
\end{eqnarray}
where $x_{f}$ is the Feynman scaling variable  and
$x_{R}$ the radial scaling variable (which depends only on the radial distance from
the kinematic boundary \cite{TA76}), with $p_{\parallel }^{\ast }$ and $p_{\parallel\max }^
{\ast }$ being the longitudinal momentum of the particle and its maximum possible value
in the center of mass ($cm$) frame respectively, while similarly $E^{\ast }$ and $E_{\max }^
{\ast }$ are the total energy of the inclusive particle and its maximum possible value
in the $cm$ frame respectively. The latter can written as $E_{\max}^{\ast }=\left( s-M_{X,
\min }^{2}+m_{p}^{2}\right) /2\sqrt{s}$, with $M_{X,\min }^{2}=2m_{p}+m_{A}/A $ being the
minimum possible mass of the recoiling particle in the considered process and $\sqrt{s}$
the total $cm$ energy. Note that for any nuclear collision, the kinematical variables used
here will always be expressed in the nucleon-nucleon ($NN$) rather than in the nucleon-nucleus
$cm$ frame, since the $NN$ $cm$ frame is the relevant physical system, the incident nucleon
energies being on the scale of 10~GeV while the average binding energy of the nucleon in
the nucleus is about 8 MeV. Bound nucleons can be considered as free particles for the 
incident protons.

A parameterization of the inclusive production cross section can be guided by some general
phenomenological features of hadron collisions (See \cite{CO84,HO73} for the general Physics
context).
\begin{itemize}
\item All experimental hadronic production cross sections show a strong exponential
decrease in transverse momentum, the exponential slope being
more or less incident energy and recoil mass $M_X$ dependent.
\item Hadronic scaling: The inclusive distribution $f\left( ab\rightarrow cX\right)$ 
of particle $c$, is to a good approximation, a function only on $p_{\perp}$ and $x_{f}$ 
(or $x_{R}$) at the high energy limit $\sqrt{s}\rightarrow \infty$. Furthermore, a large 
number of slow particles is produced (low $x_{f}$ values), the distribution decreasing 
rapidly to zero as $x_{f}\rightarrow 1$, like $\left( 1-x_{f}\right) ^{n}$. This form can 
be explained by the counting rules in parton model.

These features are predicted qualitatively by the Regge poles phenomenology and the parton model.

\item  For a given $p_{\perp}$, the inclusive distribution of produced particles is (to a good approximation for $pA$
collisions) symmetric in the rapidity space with $f(p_{\perp},y-y_{cm})=f(p_{\perp},-y+y_{cm})$, where $y_{cm}$ is 
the rapidity of the $cm$ in the laboratory frame (Lab), or in the $cm$ frame 
$f(p_{\perp},y^{\ast})=f(p_{\perp},-y^{\ast})$. A forward-backward symmetry of the cross section is expected from 
first principles (symmetry of the wave function and of the interaction) for the $NN$ system (In $pA$ collisions however, the absorption of low energy 
particles in the nuclear medium may distort the natural symmetry). In the central region where 
$y^{\ast}=0$, the inclusive distribution consists of a plateau which width increases slowly with 
the incident energy. This 
plateau reduces to a simple maximum over the energy range considered here. The inclusive
distribution rises again in the fragmentation region where $y^{\ast }\rightarrow \pm
y_{\max }^{\ast }$ for particles which can be produced diffractively but it is not the case for
$\overline{p}$ and the inclusive distribution falls simply in the fragmentation region.
\end{itemize}
%
%
\section{Parameterization of the $p+A\rightarrow \overline{p}+X$ cross section}\label{pA}
Following the approach proposed in \cite{KMN}, the former phenomenological features of
hadron collisions have been used to constrain the parameters of a functional form
describing the inclusive \pb production cross section, which could reproduce all the
relevant experimental data available from $pp$ and $pA$ collisions. The data used are listed
in Table~\ref{DATAa}. The measurements on nuclear targets cover basically the whole range
of nuclear mass, from proton to lead, over a range of incident energies from 12~GeV up to
400~GeV, matching the useful range for CR studies.

The KMN parameterization used previously \cite{KMN} is in very poor agreement with the data 
listed in Table~\ref{DATAa}, and a reexamination of the analytical approach, better 
constrained by recent data was necessary. The larger incident energy domain used here 
required some energy dependence to be introduced in the parameterization following the general 
features described above as (loose) guidelines.

In this study, the $\overline{p}$ inclusive cross section will be expressed as a function
of the three variables, $\sqrt{s}$, $p_{\perp}$ and $x_{R}$
(see for example \cite{TA76} for the relevance of the choice of these variables):
\begin{eqnarray}
\frac{Ed^{3}\sigma }{dp^{3}}=f\left( \sqrt{s},p_{\perp},x_{R}\right).
\label{eq3}
\end{eqnarray}
The following functional form used to describe the $\overline{p}$ production cross section
is an evolved version of the KMN formula:
\begin{eqnarray}
E\frac{d^{3}\sigma}{dp^{3}}&=& \sigma_{in}A^{C_{1}\ln\left(\frac{\sqrt{s}}{C_{2}}\right)p_{\perp}}
 \left(1-x_{R}\right)^{C_{3}\ln\left(\sqrt{s}\right)} e^{-C_{4}x_{R}}\nonumber \\
&&\left[C_{5}(\sqrt{s})^{C_{6}}e^{-C_{7}p_{\perp}}+C_{8}(\sqrt{s})^{C_{9}}e^{-C_{10}p_{\perp}^{2}}\right]
\label{eq4}
\end{eqnarray}
where $A$ is the target mass. The total inelastic cross section $\sigma_{in}$ for $pA$
collisions was borrowed from \cite{LE83}:
\begin{eqnarray}
\sigma _{in}\left( mb\right)  &=&\sigma _{0}[1-0.62\exp \left(
-E_{inc}/200\right) \nonumber \\
&&\sin \left( 10.9E_{inc}^{-0.28}\right) ]\nonumber \\
\sigma _{0}\left( mb\right)  &=&45A^{0.7}\left[ 1+0.016\sin \left(5.3-2.63\ln A\right) \right]
\label{eq5}
\end{eqnarray}
where $E_{inc}$ is the incident kinetic energy in MeV.

The 10 parameters $C_{1}$-$C_{10}$ have been fitted to the set of experimental data listed
in Table~\ref{DATAa} by a standard $\chi ^{2}$ minimization procedure using the code
\textit{MINUIT} \cite{MINUIT}.

In relation (\ref{eq4}), the term $(1-x_{R})^{C_{3}}$ originates from the hadronic scaling
properties, namely, the quark counting rules of the parton model of hadronic interactions
\cite{CO84} (see Sect.~\ref{FEAT}). It was found empirically in this study, that a
significantly better result is obtain if the exponent is energy dependent. The
$\mbox{ln}(\sqrt{s})$ factor multiplying the $C_{3}$ coefficient was found to give the
best result.
The term $e^{-C_{4}x_{R}}$ is induced by the Regge regime \cite{CO84}. The last factor of
relation~(\ref{eq4}) accounts for the transverse momentum dependence of the cross section
(see Sect.~\ref{FEAT}). The analysis of the experimental data (Table~\ref{DATAa})
showed that the term of angular dependence $e^{-C_{10}p_{\perp}^{2}}$ is dominant at
low energy, $E_{p}^{lab}\approx 10~$GeV, while the term $e^{-C_{7}p_{\perp}}$ dominates
at high energies, $E_{p}^{lab} > 100~$GeV.
The $\sqrt{s}$ dependence has been introduced to allow the transition from the $p_{\perp}$
to the $p_{\perp}^{2}$ dependence. The target mass
dependence was accounted for by the factor $A^{C_{1}\ln \left( \frac{\sqrt{s}}{C_{2}}
\right) p_{\perp }}$, with an energy dependent exponent introduced for the same reason as
above, at variance with the constant exponent used in the KMN parameterization.
The energy dependence used (linear in $\sqrt{s}$) accounts for the experimental increase 
of this coefficient with the incident energy found using the KMN approach. For
incident energies $E_{p}^{lab} < 55~$GeV, this coefficient becomes negative.
%
\begin{figure}[hbtp!]
\includegraphics[width=9cm]{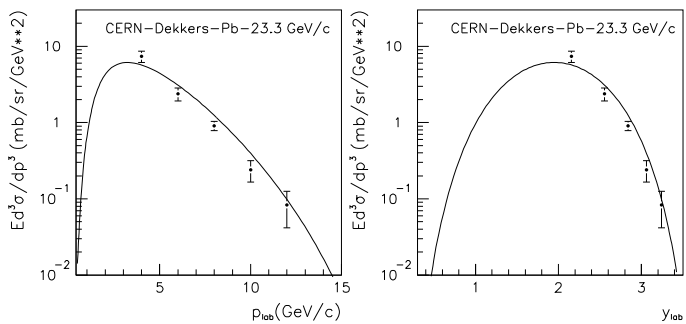}
\caption{\label{figure1}
\small Experimental \pb production cross section for $p+Pb$ collisions at 23.1 GeV/$c$
and at 0 degree scattering angle
\cite{DE65} compared with the parameterization (\ref{eq4}), plotted against particle
momentum $p_{lab}$ (left), and rapidity $y$ (right). In this latter case the distribution
is symmetric around the value of the $cm$ rapidity $y_{cm}=1.9$, corresponding to
$p_0\sim 3.3$ GeV/$c$ in the Lab. The fit to the cross section data in the upper $p>p_0$
region of the momentum region also determines the values of the cross section for  $p<p_0$
where no data is available.}
\end{figure}

In Sect. 2, it was mentioned that one of the features of the inclusive distribution is its 
symmetry in the rapidity space. By construction our parameterization (relation 
(\ref{eq4})) satisfies this symmetry property since it depends only on $\sqrt{s}$, 
$p_{\perp}$ and $x_{R}$. This is illustrated in Fig.\ref{figure1} which shows the fit 
to the 23.1~GeV/$c$ cross section data versus particle rapidity (right) and particle 
momentum (left). The maximum of the cross section corresponds to particle production
with velocity zero in the $cm$ frame, i.e., traveling with the $cm$ velocity in the laboratory. 
The corresponding value of this momentum in the Lab is: $p_{0}\approx\sqrt{m_p E_p/2}$. 
Since the cross section distribution is symmetric in the rapidity space, the upper branch 
of rapidity with respect to the $y_{cm}$, completely determines the values of cross 
section along the lower branch. This is true as well in the Lab frame where the values of
the cross section for $p>p_{0}$ determine the values below this momentum. In the case of 
the figure this means that a fit to the experimental values of the cross section above 
$p_{0}\approx 3.3$~GeV/$c$ also determines the values of the cross section below with
the same level of accuracy. Fitting the data from about $y_{lab}\sim$~2.1 up to 
$y_{lab}\sim$~3.2, also determines the cross section down to $y_{lab}\sim$~0.65, i.e., 
down to $p_{lab}\sim$~0.65~GeV/c, the validity of the fit extending likely significantly 
below this value. This is an important point since experimental \pb cross section data are 
usually scarce below about 1~GeV, and since the production cross section for these low energy
particles is very important because \pbs originating from neutralino (dark-matter) 
annihilation \cite{JU94} or from the evaporation of primordial black holes \cite{BA00}, are 
expected within this energy range.
%
%
\begin{figure}[hbtp!] %
\includegraphics[width=9cm]{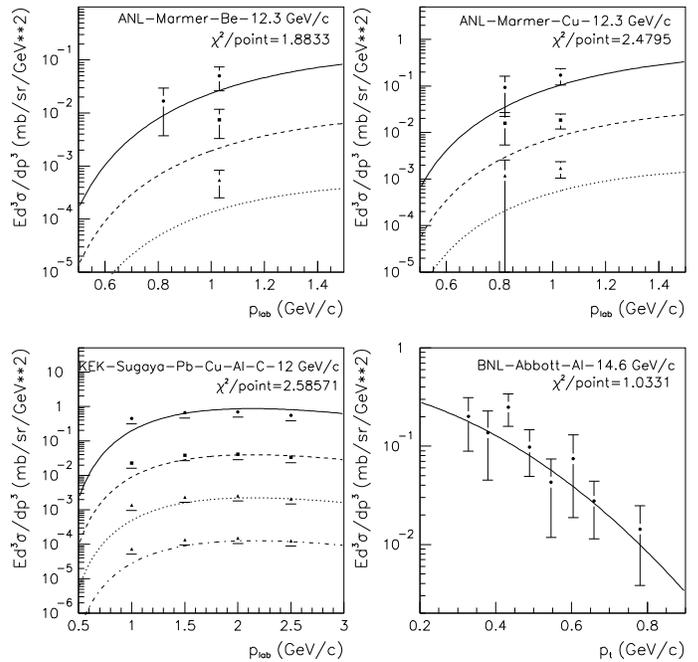}
\caption{\small Experimental data points from \cite{SU98, MA69, AB93} compared to the best fit calculations.
\label{figure2}}
\end{figure}
\begin{figure}[hbtp!] 
\includegraphics[width=9cm]{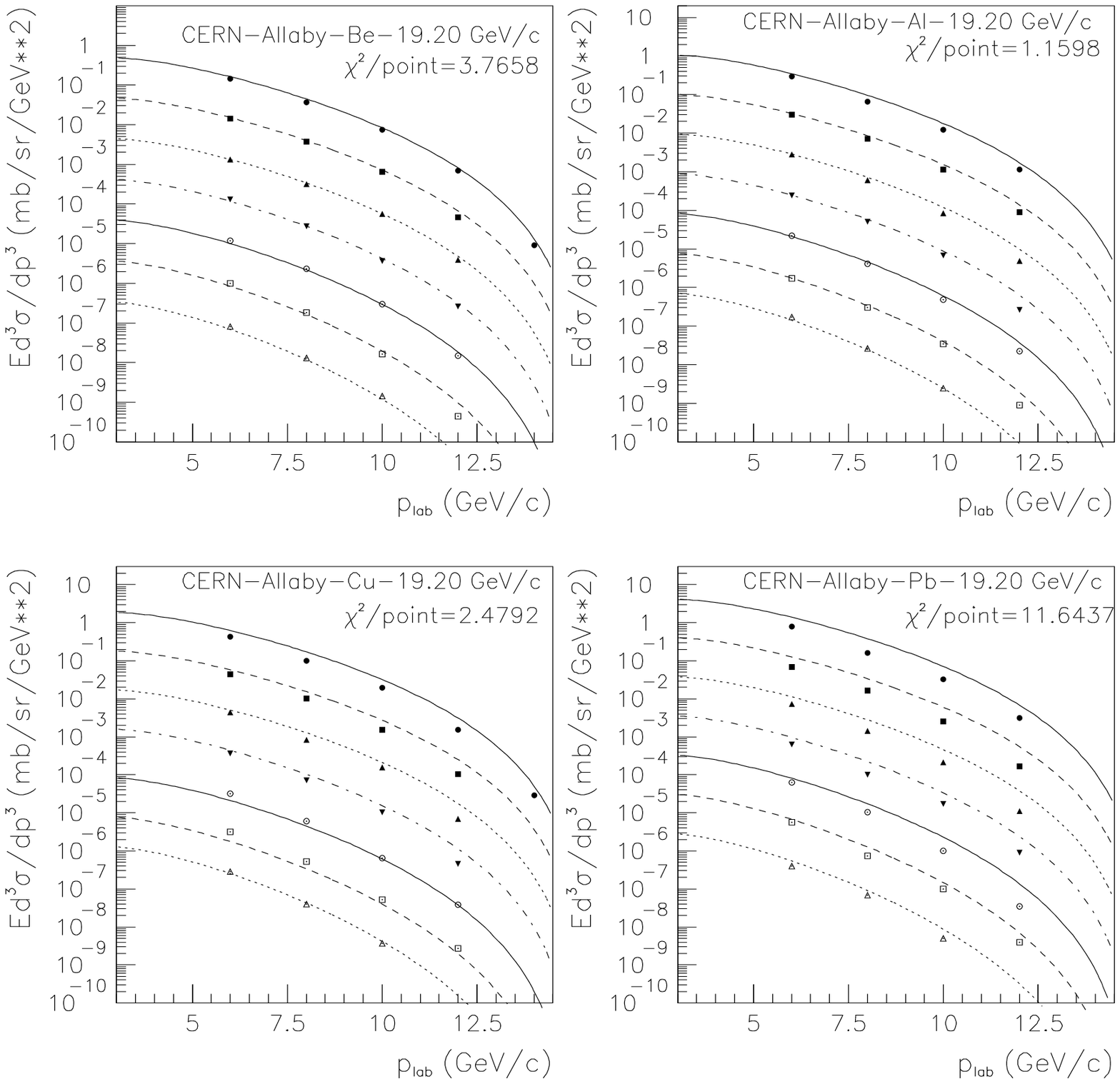}
\caption{\small Experimental data points from \cite{AL70} compared to the best fit calculations.
\label{figure3}}
\end{figure}
\begin{figure}[hbtp!] 
\includegraphics[width=9cm]{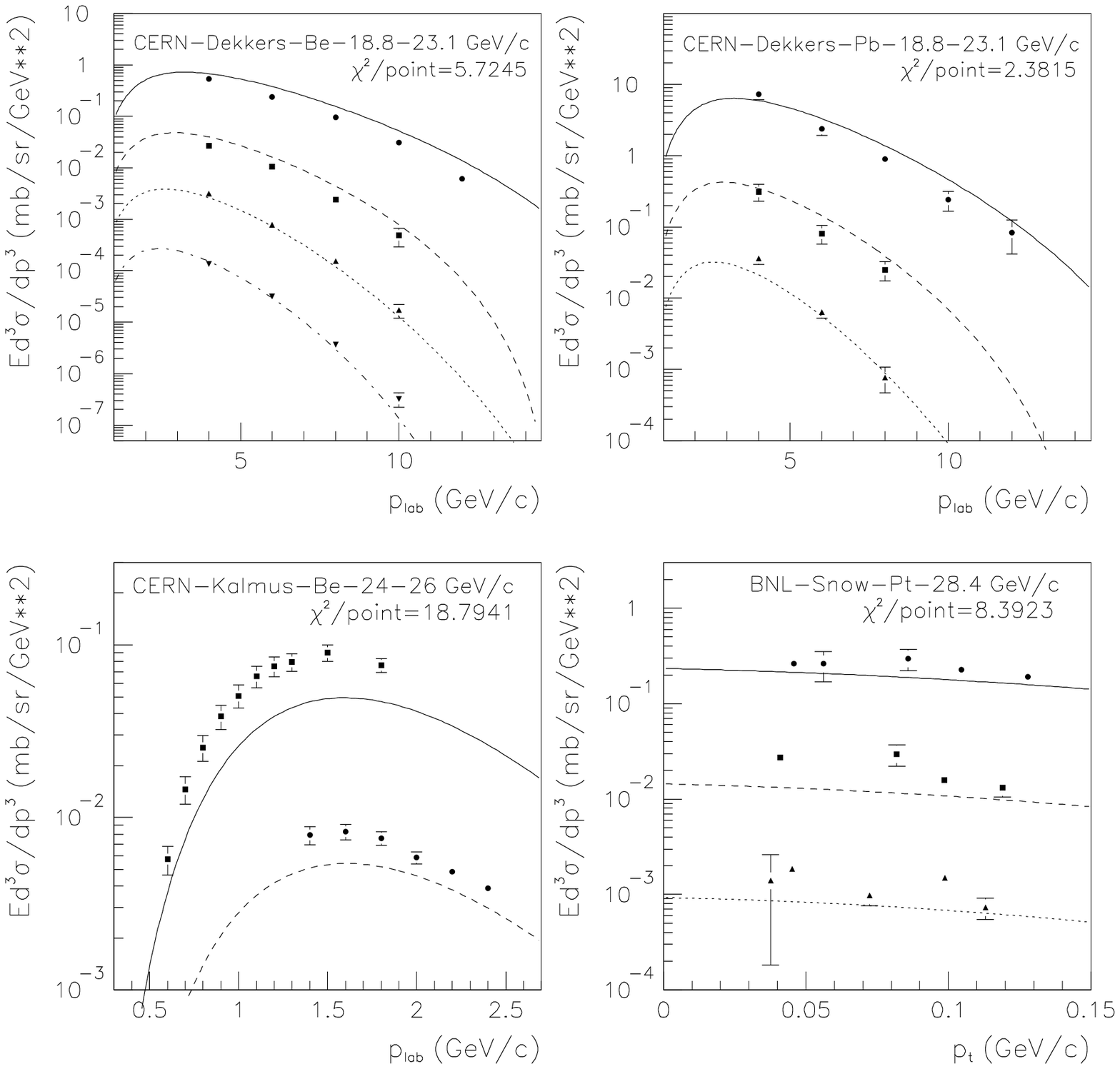}
\caption{\small Experimental data points from \cite{DE65, KA71, SN85} compared to the best fit calculations.
\label{figure4}}
\end{figure}
\begin{figure}[hbtp!] 
\includegraphics[width=9cm]{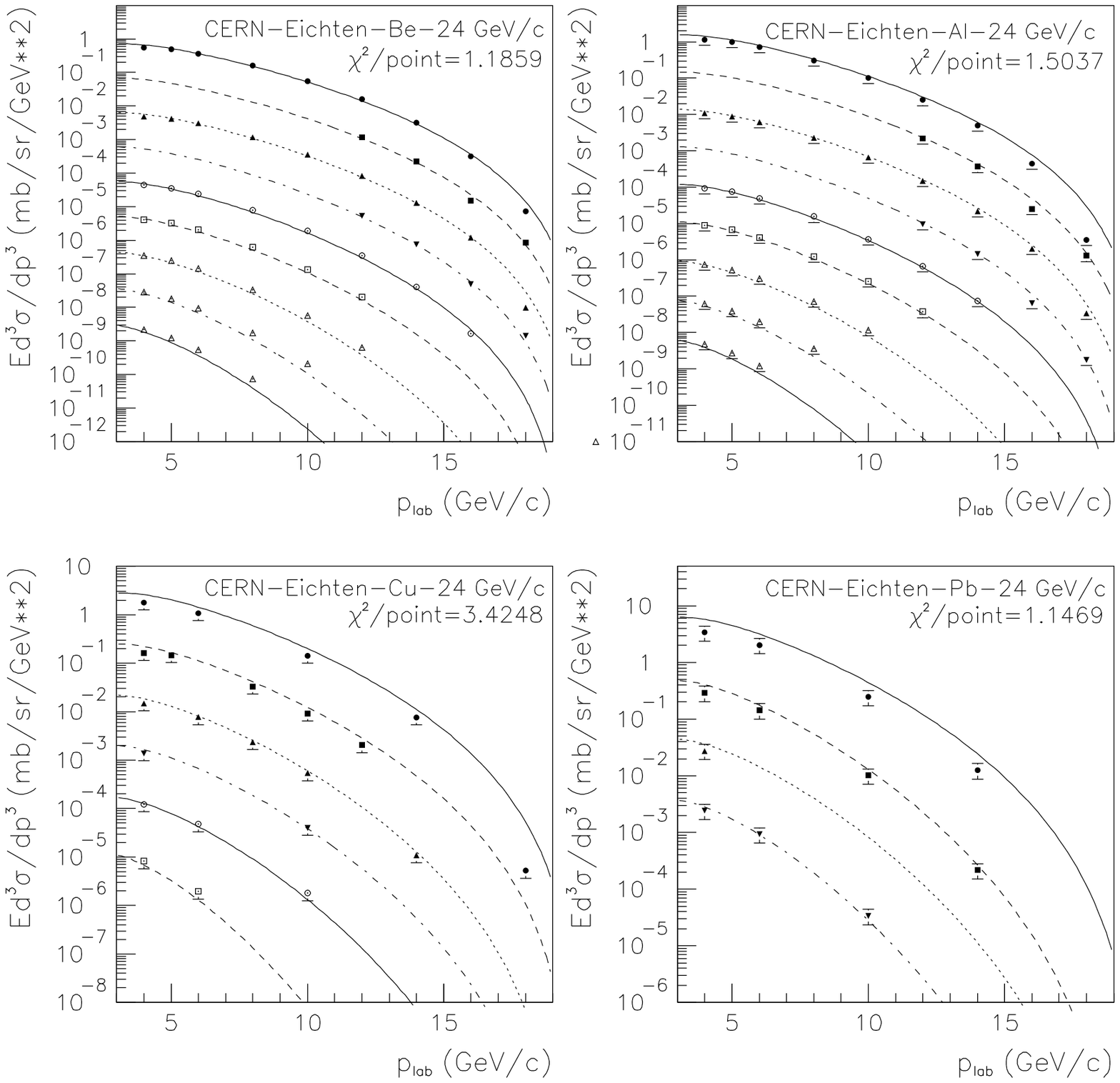}
\caption{\small Experimental data points from \cite{EI72} compared to the best fit calculations.
\label{figure5}}
\end{figure}
\begin{figure}[hbtp!] 
\includegraphics[width=9cm]{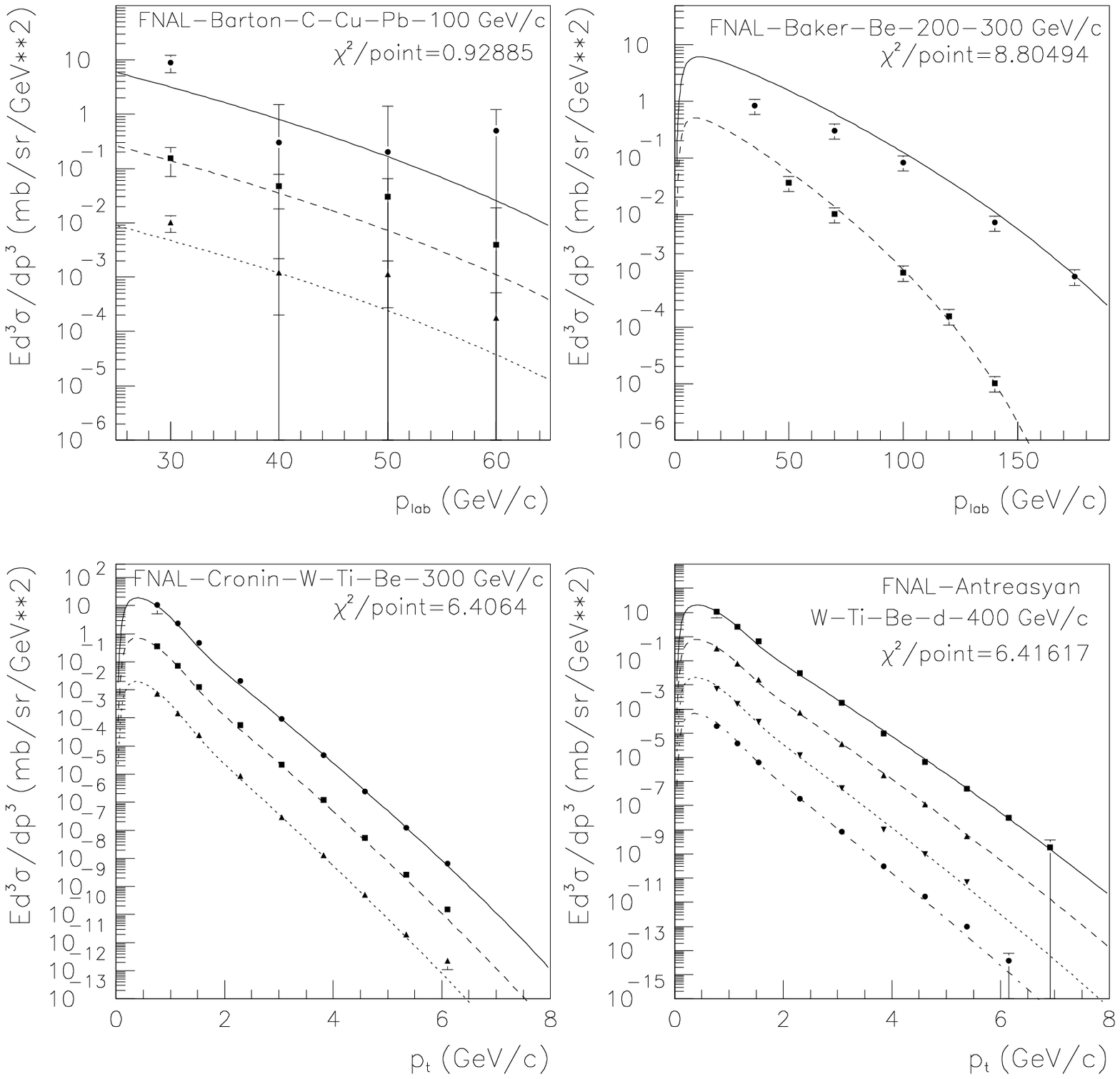}
\caption{\small Experimental data points from \cite{BA83, BA74, CR75, AN79} compared to the best fit calculations.
\label{figure6}}
\end{figure}
%
\begin{table*}
\begin{center}
\begin{tabular}{|c|l|l|l|l|}\hline\hline
Experience & Target & $p_{inc}$ or $\sqrt{s}$& \pb kinematical range & $\theta_{lab}$  \\
           &        & GeV/$c$ (GeV)              &    GeV/$c$            & mrad
\\ \hline\hline
Y.Sugaya \textit{et al} & C, Cu, Al, Pb &12& $p_{lab}$: 1.0$-$2.5 & 89  \\
KEK-PS 1998 \cite{SU98} &&&& \\ 
G.J.Marmer \textit{et al} & Be, Cu & 12.3 & $p_{lab}$: 0.820, 1.030 & 0, 0.17, 10  \\
ANL 1969 \cite{MA69} &&&& \\ 
T.Abbott \textit{et al} & Al &14.6& $m_{t}$: 0$-$0.3 &    \\
BNL 1993 \cite{AB93} &&&$y$: 1.0$-$1.6& \\ 
J.V.Allaby \textit{et al} & $p$, Be, Al, Cu, Pb & 19.20 & $p_{lab}$: 4.5$-$18.3 & 12.5$-$70  \\
CERN 1970 \cite{AL70} &&&& \\ 
D.Dekker \textit{et al} & $p$, Be, Pb & 18.8, 23.1 & $p_{lab}$: 4$-$12 & 0, 100  \\
CERN 1965 \cite{DE65} &&&& \\ 
T.Eichten \textit{et al} & Be, Al, Cu, Pb & 24 & $p_{lab}$: 4$-$18 & 17-127  \\
CERN 1972 \cite{EI72} &&&& \\ 
P.I.P.Kalmus \textit{et al} & Be & 24$-$26 & $p_{lab}$: 0.6$-$2.5 & 310  \\
CERN 1971 \cite{KA71} &&&& \\ 
J.M.Snow \textit{et al} & Pt & 28.4& $p_{lab}$: 0.606$-$0.730 & 0$-$0.17   \\
BNL 1985 \cite{SN85} &&&& \\ 
D.S.Barton \textit{et al} & $p$, C, Cu, Al, Ag, Pb & 100& $p_{lab}$: 30$-$88 &   \\
FNAL 1983 \cite{BA83} &&&$p_{\perp}$: 0.3, 0.5& \\ 
J.R.Johnson \textit{et al} & $p$ & 100, 200, 300 & $p_{\perp}$: 0.25$-$1.5 &   \\
FNAL 1978 \cite{JO78} &&&$0.05<x_{R}<1.0$& \\ 
W.F.Baker \textit{et al} & Be & 200, 300 & $p_{lab}$: 23$-$197 & 3.6  \\
FNAL 1974 \cite{BA74} &&&& \\ 
J.W.Cronin \textit{et al} & Be, Ti, W & 300 & $p_{\perp}$: 0.76$-$6.91 & 77  \\
FNAL 1975 \cite{CR75} &&&& \\ 
D.Antreasyan \textit{et al} & $p$, $d$, Be, Ti, W & 200, 300, 400 & $p_{\perp}$: 0.77$-$6.91 &  $77$ \\
FNAL 1979 \cite{AN79} & &&& \\ 
K.Guettler \textit{et al} & $p$ & $23<\sqrt{s}<63$ & $p_{\perp}$: 0.1$-$0.3 &   \\
CERN 1976 \cite{GU76} &&&$x_{f}=0$& \\ 
P.Capiluppi \textit{et al} & $p$ & $23.3<\sqrt{s}<63.7$ & $p_{lab}$: 1.5$-$10& 80$-$350  \\
CERN 1974 \cite{CA74} &&&& \\ \hline\hline
\end{tabular}
\small
\caption{\small List of the experimental $\overline{p}$ production cross section data
included in the $\chi ^{2}$ minimization procedure, in increasing energy order.
\label{DATAa}}
\normalsize
\end{center}
\end{table*}

\section{Results for nuclear targets}\label{RES}
The data used in the fit procedure are summarized in Table~\ref{DATAa}. The fit sample
included measurements from 12~GeV up to 400~GeV incident proton Lab energy on nuclear
targets going from Deuterium up to Pb nuclei, and for momentum transfers up to 6.91~GeV/$c$.
For $pp$ collisions, the incident $cm$ energy $\sqrt{s}$ extended from about 6 up to
63~GeV.
%
\begin{table*}
\begin{center}
\begin{tabular}{|c||c|c|c|c|c|}\hline
parameter & $C_{1}$ & $C_{2}$ & $C_{3}$ & $C_{4}$ & $C_{5}$ \\ \hline
value(error)  &   0.16990(4)  &  10.28(13)  &  2.269(7)  &   3.707(27) &   0.009205(2)\\ \hline
parameter & $C_{6}$ & $C_{7}$ & $C_{8}$ & $C_{9}$ & $C_{10}$ \\ \hline
value(error)  &   0.4812(14)  &  3.3600(2)  &  0.063940(73)  &  $-$0.1824(15)&   2.4850(6)\\ \hline
\end{tabular}
\small
\caption{\small Values of the parameters of relation (\ref{eq4}) obtained from fitting
the experimental $\overline{p}$ production cross sections list in Table~\ref{DATAa} and 
the corresponding error following the PDG standard conventions.
\label{PARA}}
\normalsize
\end{center}
\end{table*}
\begin{table}
\begin{center}
\begin{tabular}{|c|c|c|}\hline
system   & parameterization            & $\chi ^{2}$ per point \\ \hline \hline
   $pp,pA$  & KMN \cite{KMN}         &   80.0   \\  \hline
   $pp,pA$  & this work (\ref{eq4})             &   5.3    \\  \hline \hline
   $pp$    & Tan and Ng \cite{TA82}  &   28.1   \\  \hline
   $pp$    &  this work (\ref{eq6})            &   3.6   \\  \hline
\end{tabular}
\small
\caption{\small Comparison between this work and the other parameterizations.
\label{CHI2}}
\normalsize
\end{center}
\end{table}

The $\chi^{2}$ per point obtained with the parameterized relation (\ref{eq4}) is 5.32 
(Table~\ref{CHI2}) for 654 experimental points (see list in Table~\ref{DATAa}). The values 
of the parameters obtained in the fit are given in Table~\ref{PARA} together with error. 
The correlation coefficients between the parameters determined in the search are given in appendix. The results for 
nuclear targets (A$\geq$2) are shown on Figs~\ref{figure2} to \ref{figure6} where the data 
points are compared with the calculated values are given in appendix. The values of the  $\chi^{2}$ per point of 
each set are given in the figures. In each case, some basic informations (Authors, beam 
energy, target nuclei, and $\chi ^{2}$ per point obtained for the considered set) are given 
on the figures. In all the figures, the top distribution corresponds to the measured cross 
section, while each next distribution below has been multiplied by a 10$^{-1}$ factor with 
respect to the previous one, for the legibility of the figure.

As it can be seen on the figures, the quality of the fits varies from fair to excellent. A 
poor fit is obtained however for the 24--26 GeV data from \cite{KA71}, the calculations 
underestimating the data by a factor of about 2. 
Nevertheless, this set has been kept in the fit procedure since its contribution is small 
and not hardly affects the results (which is not the case for the data listed in
Table~\ref{DATAb}). To the opposite, outstandingly good fits have been obtained consistently 
and simultaneously for the CERN data from \cite{AL70,EI72} in the 20--25 GeV incident energy 
range, and for the high energy and large momentum transfer data from \cite{CR75,AN79}.

Table~\ref{CHI2} compares the values of the  $\chi^{2}$ per point obtained in the present 
study, with that obtained using the KMN relation \cite{KMN} for the same data. The latter is 
seen to be more than one order of magnitude larger than the value obtained using (\ref{eq4}). 
This gives the scale of the improvements achieved by the present study on the issue.

These results demonstrate the ability of the proposed parameterization to describe the 
inclusive \pb production cross section on nuclei over the quoted ranges of incident energy,
momentum transfer, and target mass, with a good accuracy.
%
\begin{table*}
\begin{center}
\begin{tabular}{|c|c|c|c|c|}\hline\hline
Experience & target & $p_{inc}$/$E_{inc}$  & \pb kinematical range &
$\theta_{lab}$  \\ &&(GeV/$c$)&(GeV/$c$)&(mrad)
\\ \hline \hline
Yu.M.Antipov \textit{et al} & Al & 70 & $p_{lab}$: 10$-$60 &0  \\
IHEP 1971\cite{AN71} &&&& \\ 
V.V.A.Abramov \textit{et al} & C, Al, Cu, Sn, Pb & 70 & $p_{\perp}$: 0.99$-$4.65 &160  \\
IHEP 1984 \cite{AB84} & &&& \\ 
W.Bozzoli \textit{et al} & Be, Al, Pb & 200 & $p_{lab}$: 20$-$37 &0  \\
CERN 1978 \cite{BO78} & &&& \\ 
I.G.Bearden \textit{et al} & Be, S, Pb & 450 & $p_{lab}$: 4$-$8.5 &37, 131  \\
CERN 1998 \cite{BE98} &$$ &&$p_{\perp}$: 0.11$-$1.28& \\ \hline\hline
\end{tabular}
\small
\caption{\small Antiproton production cross section data not taken into account in the
$\chi ^{2}$ minimization procedure, classified by increasing energy. See text for explanations.
\label{DATAb}}
\normalsize
\end{center}
\end{table*}
%
\subsection{Data discarded from the selection}\label{REJ}
%
\begin{figure}[hbtp!]
\vspace{-0.5cm}
\includegraphics[width=9cm]{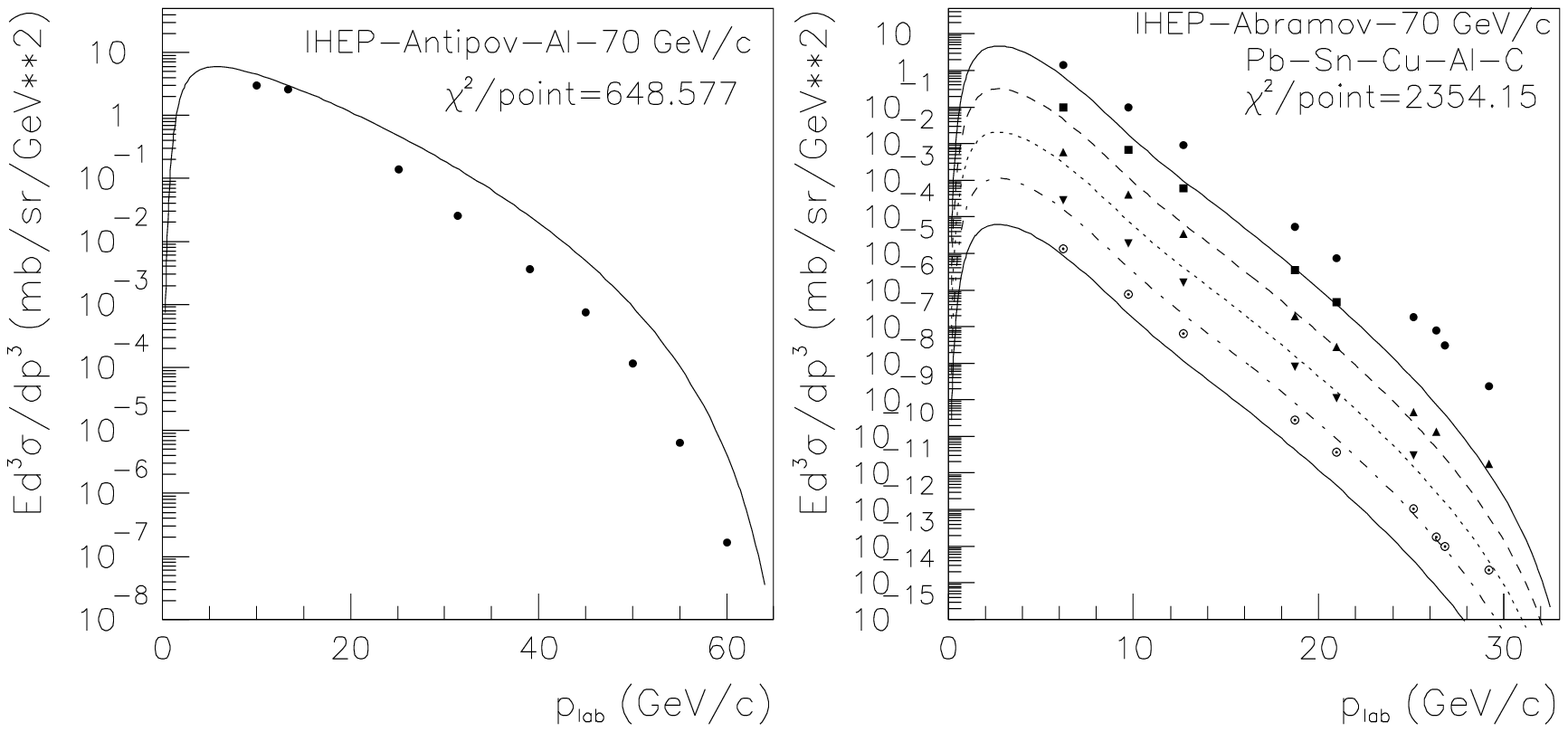}
\vspace{-1cm}
\caption{\small Experimental data points from \cite{AN71,AB84} compared to the best fit
calculations. Data points not included in the final search procedure.
\label{figure7}}
\end{figure}
%
\begin{figure}[hbtp!]
\vspace{-0.5cm}
\includegraphics[width=9cm]{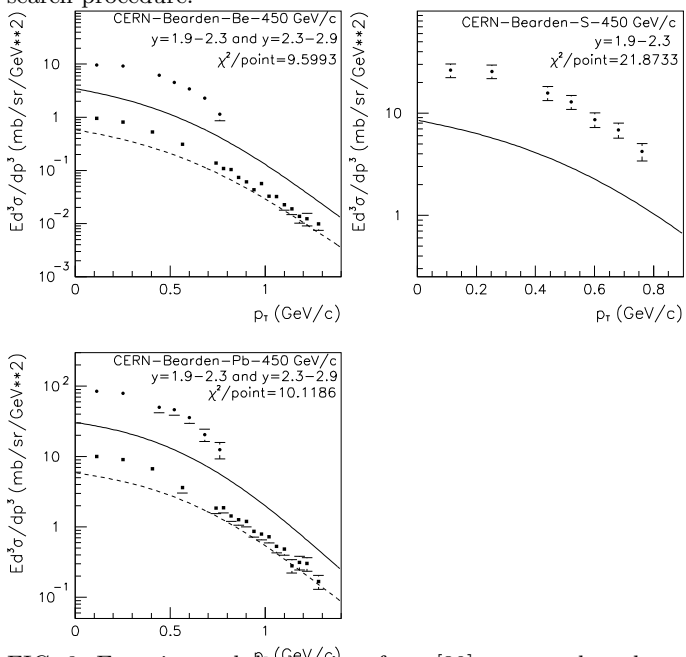}
\vspace{-1cm}
\caption{\small Experimental data points from \cite{BE98} compared to the best fit calculations. Data points not included in the final search procedure.
\label{figure8}}
\end{figure}
%
The data listed in Table~\ref{DATAb} were not included in the fit sample, because of their
obvious inconsistency with the other data. This is illustrated on Figs.~\ref{figure7} and
\ref{figure8} where they are compared to the best fit calculations obtained in the
previous step on the selected sample.
As it can be seen, the difference  between data and calculated values amounts up to about
one order of magnitude. The ratio goes from 2 to 10 for the Serpukhov experiments 
\cite{AB84,AN71}.  For \cite{AN71}, it is about 5, and consistent with a simple normalization 
problem.

A larger and more surprising disagreement is found with some recent CERN data from NA44
\cite{BE98}, in particular for the measurements in the small rapidity bin. Note also
that the parameterization (\ref{eq6}) describes quite well the data from \cite{AN79} obtained on 
the same targets as \cite{BE98} over a wider kinematical region (see Fig.~\ref{figure6}).

The \pb cross section data from \cite{BO78} appearing in the table, were given in the original 
works in units of the corresponding $\pi^-$ production cross section measured at the same 
momentum. Although the absolute value could be obtained using the known $\pi^-$ cross section, 
the results were considered too inaccurate however, and discarded from the selected sample.
%
\subsection{Analysis of the $pp\rightarrow\overline{p}X$ data}\label{PP}
%
\begin{table*}
\begin{center}
\begin{tabular}{|c||c|c|c|c|c|c|c|}\hline
parameter & $D_{1}$ & $D_{2}$ & $D_{3}$ & $D_{4}$ & $D_{5}$ & $D_{6}$ & $D_{7}$\\ \hline
value(error)     &   3.4610(20) &  4.340(20) &  0.007855(3) &  0.5121(27) &   3.6620(5) &   0.023070(1) &   3.2540(77)\\ \hline
\end{tabular}
\small
\caption{\small Values of the parameters of relation (\ref{eq6}) obtained
by fitting the experimental $\overline{p}$ production cross sections list in Table 3 for proton-proton collisions and the corresponding error
following the PDG standard conventions.
\label{PARA2}}
\normalsize
\end{center}
\end{table*}
%
\begin{figure}[hbtp!]
\includegraphics[width=9cm]{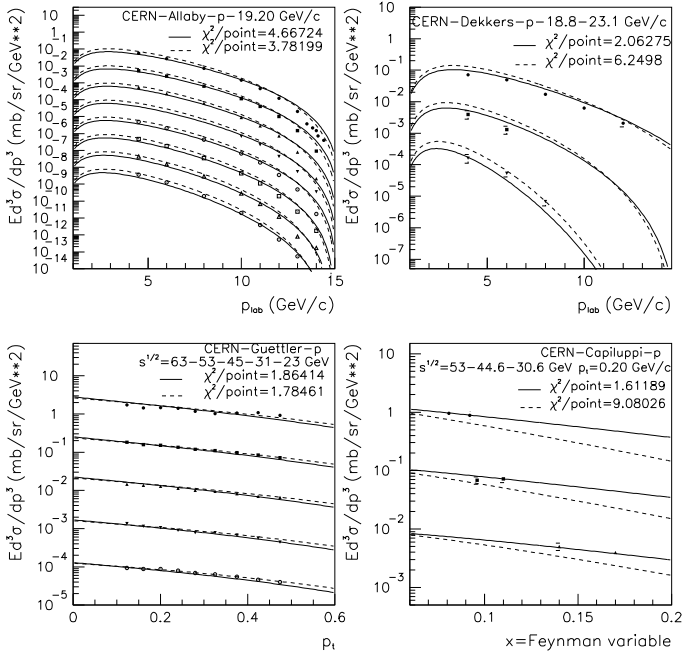}
\caption{\small Experimental data points from pp collisions from
\cite{AL70, DE65, GU76, CA74} compared to the best fit calculations using the
two parameterizations. See text. \label{figure9}}
\end{figure}
\begin{figure}[hbtp!]
\includegraphics[width=9cm]{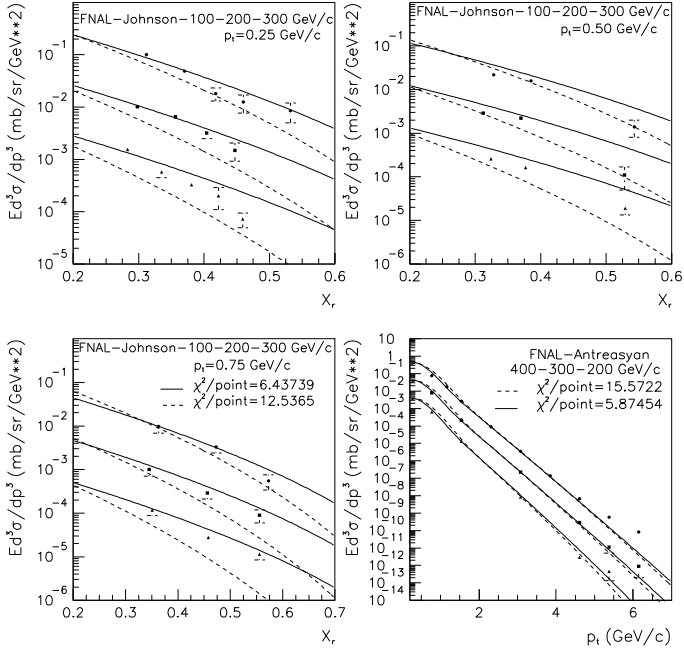}
\caption{\small Same as Fig.~\ref{figure9} for the data from \cite{JO78, AN79},
The $\chi^{2}$ per point for the first three graph is indicated in graph 3.
\label{figure10}}
\end{figure}
%
This reaction is the dominant contribution to the secondary $\overline{p}$ production
induced by Cosmic Rays, since the interstellar gas is mainly constituted of hydrogen gas.
It is thus important to obtain as accurate a description as possible for the cross section.

Considering separately the $p+p$ collision data in Table~\ref{DATAa}, the parameterization
(\ref{eq4})  gives for the best fit a value of the $\chi^{2}$
per point of 7.08. For the same data, the well known parameterization of Tan and Ng
\cite{TA82} gives a value of 28.1. In addition, this latter parameterization is valid only
for $p_{\perp}=0-0.8$ GeV/$c$ and is not able to reproduce the large $p_{\perp}$ data such
as those from \cite{CR75} and \cite{AN79} where $p_{\perp}=0.76-6.91~$GeV/$c$.
Note also that Tan and Ng's parameterization contains 8 parameters for
$\sqrt{s}>10$~GeV $(p_{lab}>50$~GeV/$c$) and 17 for $\sqrt{s}<10$~GeV.

However, in the course of the study, it appeared that some of the parameters of
relation~(\ref{eq4}) had no incidence on the resulting fits.
The parameterization (\ref{eq4}) has then been revisited and simplified from some of its
parameters irrelevant for this particular reaction and from other parameters which turned
out to be ineffective in the minimization procedure, resulting in the following functional
form for $pp\rightarrow\overline{p}X$ inclusive production cross section :
\begin{eqnarray}
E\frac{d^{3}\sigma}{dp^{3}}&=& \sigma_{in}
 \left(1-x_{R}\right)^{D_{1}} e^{-D_{2}x_{R}}\nonumber \\
&&\left[D_{3}(\sqrt{s})^{D_{4}}e^{-D_{5}p_{\perp}}+D_{6}e^{-D_{7}p_{\perp}^{2}}\right]
\label{eq6}
\end{eqnarray}
In comparison with relation (\ref{eq4}), the dependence with the mass of the target has been
removed since the only proton target is considered in this case. In addition, the energy
dependent factors $\sqrt{s}$ in front of $D_{1}$ and $D_{7}$ in (\ref{eq4}) have been also
removed because of their ineffectiveness in the minimization procedure.

The parameters $D_{1}$ to $D_{7}$ have been adjusted by the same $\chi^{2}$ minimization
procedure as previously \cite{MINUIT}, to the set of experimental data listed in
Table~\ref{DATAa} restricted to $pp$ collisions. With formula (\ref{eq6}), the $\chi^{2}$
per point obtained for the best fit is 3.59, for 228 experimental point, instead of 7.08
with relation (\ref{eq4}). The values of the fit parameters obtained with (\ref{eq6}) 
are given in Table~\ref{PARA2}. The correlation coefficients between the parameters
determined in the search are given in appendix.
Note that the values of the coefficients $C_{3}$ and $D_{2}$, $C_{5}$ and $D_{3}$,
respectively, are of the same order of magnitude. This was expected since they describe the same physics
in the relations (\ref{eq4}) and (\ref{eq6}).

Figs.~\ref{figure9} and \ref{figure10} show the $pp\rightarrow\overline{p}X$ data 
analyzed, compared with the best fit results obtained for the whole $pp$ and $pA$ data 
sets from Table~\ref{DATAa}, using relation~(\ref{eq4}) (dashed line) and with those 
obtained for the $pp$ data only using relation~(\ref{eq6}) (solid line). The simplified 
form (\ref{eq6}) clearly provides a significantly better account of the measured cross 
sections, the $\chi^{2}$ value obtained being better by a factor of about 2 (about 3.6 
against about 7).

The results obtained in this work have been used in the calculations of the $\overline{d}$,
$\overline{t}$ and $\overline{\mbox{He}}$ production from $p+p$ and $p+A$ collisions in
the atmosphere and in the galaxy \cite{DU03a,DU03b,DU03c}.
%
%
\section{Antiproton mean multiplicity}
In this section, the antiproton mean multiplicity, defined as
\begin{eqnarray}
\left\langle n_{\overline{p}}\right\rangle =\frac{1}{\sigma _{in}}\int f%
\frac{d^{3}p}{E},
\end{eqnarray}
and depending only on $\sqrt{s}$, has been computed by means of relations (\ref{eq4}) and 
(\ref{eq6}) and compared with the experimental data in $pp$ collision \cite{AN73}.
Note that the original data from \cite{AN73} have been corrected from the single-diffractive 
contribution to the total inelastic cross section $\sigma _{in}$ \cite{GI01}. The corrected 
antiproton mean multiplicity should thus be somewhat smaller than the measured values 
(by $\sim$15--20\%).
%
\begin{figure}[hbtp!]
\includegraphics[width=9cm]{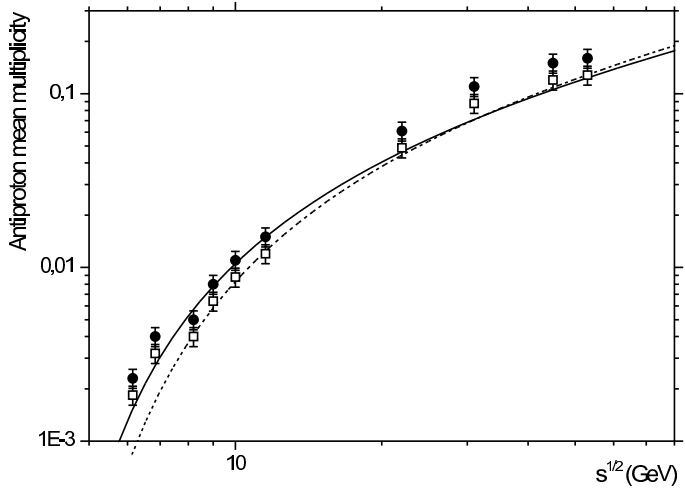}
\caption{\small Antiproton mean multiplicity distribution in the whole phase space,
calculated using relation (\ref{eq4}) (solid line), and relation (\ref{eq6}) (dashed line),
compared with experimental data \cite{AN73}, uncorrected (full circles) and corrected (open 
squares) from diffractive contribution. See text for details. \label{figure11}}
\end{figure}

The results, shown on Fig.~\ref{figure11}, are in good agreement with the experimental data.
Note that below $\sqrt{s}\,\simeq$~15 GeV/$c$, the results given by the relation
(\ref{eq4}) and (\ref{eq6}) become significantly different (about a factor 2 at the maximum).
As expected, the simplified form (\ref{eq6}) gives a little bit better results
since the experimental data are from $pp$ collisions.
%
\section{Conclusion}
The parameterization of the \pb inclusive production cross section on nuclei has been 
revisited by investigating a broad collection of data sets available, covering a large 
dynamical domain of incident energy and of momentum transfer, for a broad range of nuclear 
masses. Good results have been obtained but for a small sample of data sets inconsistent with 
the other data. The experimental \pb inclusive production cross sections can be reproduced to 
within a few tens of percent over this range, i.e., for incident energies from 12~GeV up to 
400~GeV, and for target mass $1\leq A\leq 208$. Theses results constitute a significant 
improvement with respect to the former KMN parameterization, decreasing by a factor of about 
15 the value of the $\chi^{2}$ per point obtained using the latter. A simplified version of the
functional form has been developed for $pp$ collisions giving also good results up to very high 
energies, much beyond the range dictated by the Cosmic Ray Physics requirements which motivated
the study. This also constitutes an improvement, consistent with the data on nuclei, of the Tan 
and Ng formula used so far as a standard in the calculations.

The parameterisations (\ref{eq4}) and (\ref{eq6}) are also able to reproduce the experimental 
antiproton mean multiplicity measured in $pp$ collision with a good accuracy.

A point to be emphasized is that because of the symmetry of the cross section in the rapidity
space, the fitted range in the Lab momentum of the particle, usually measured above the
$cm$ rapidity, also determines the cross section at low momenta, a range of major importance
for Cosmic Ray antiprotons where accuracy is extremely important.

\acknowledgements
The authors are grateful to Michael Murray for helpful discussions on the NA44 data.

\section{Appendix}
The symmetrical matrix (\ref{eq8}) and (\ref{eq9}) give respectively the correlation coefficients
for the parameters $C_{1}$$-$$C_{10}$ and $D_{1}$$-$$D_{7}$ of equations (\ref{eq4})
and (\ref{eq6}) respectively.

In relation (\ref{eq4}), the coefficients $C_{1}$ and $C_{2}$ appear to be strongly 
correlated (correlation coefficient 0.961), as it could be expected from their functional 
dependence. On the contrary, coefficient $C_{5}$ and $C_{8}$ are not correlated (correlation 
coefficient 0.232), since effective in different energy ranges (see Sect.~\ref{pA}).
The same remarks apply to the coefficients $D_{1}$$-$$D_{7}$.
\begin{widetext}
\begin{equation}
\left(
\begin{array}{rrrrrrrrrr}
1.000 & 0.961 & 0.120 & -0.200 & -0.148 &0.128 & 0.086 & -0.067 & -0.048 &
-0.165 \\
& 1.000 & 0.131 & -0.209 & -0.199 & 0.157 & -0.098 & -0.067 & -0.060 & -0.148 \\
&  & 1.000 & -0.937 & -0.321 & 0.228 & -0.049 & -0.655 & -0.620 & -0.289 \\
&  &  & 1.000 & 0.282 & -0.180 & -0.042 & 0.834 & 0.784 & 0.311 \\
&  &  &  & 1.000 & -0.962 & 0.358 & -0.110 & -0.128 & 0.239 \\
&  &  &  &  & 1.000 & -0.164 & 0.232 & 0.286 & -0.202\\
&  &  &  &  &  & 1.000 & -0.127 & 0.007 & -0.028 \\
&  &  &  &  &  &  & 1.000 & 0.979 & 0.210 \\
&  &  &  &  &  &  &  & 1.000 & 0.148 \\
&  &  &  &  &  &  &  &  & 1.000
\end{array}
\right)
\label{eq8}
\end{equation}
\begin{center}
\small
Correlation coefficients for the parameters $D_{1}$$-$$D_{10}$ given in
Tables~\ref{PARA}.
\normalsize
\end{center}
\end{widetext}
\begin{widetext}
\begin{equation}
\left(
\begin{array}{rrrrrrr}
1.000 & -0.933 & -0.557 & 0.604 & 0.080 & -0.311 & 0.087 \\
& 1.000 & 0.654 & -0.730 & -0.141 & 0.435 & -0.135 \\
&  & 1.000 & -0.979 & 0.502 & -0.212 & -0.241 \\
&  &  & 1.000 & -0.336& 0.033 & 0.228 \\
&  &  &  & 1.000 & -0.833 & -0.213 \\
&  &  &  &  & 1.000 & 0.282 \\
&  &  &  &  &  & 1.000
\end{array}
\right)
\label{eq9}
\end{equation}
\begin{center}
\small
Correlation coefficients for the parameters $C_{1}$$-$$C_{7}$ given in
Tables~\ref{PARA2}.
\normalsize
\end{center}
\end{widetext}
%
%


\begin{thebibliography}{$NN$}
%
\bibitem{JU94} see for example  G. Jungman \textit{et al}., {\ Phys. Rev. D.}
{\bf 49}, 2316 (1994).
\bibitem{BA00} A. Barrau \textit{et al}., A\&A {\bf 288}, 676 (2002)
\bibitem{KMN} A.N. Kalinovskii, M.V. Mokhov, and Yu.P. Nikitin,
              \textit{Passage of high-energy particules through matter}, American Institute
              of Physics, 1989.
\bibitem{HU02} C.Y. Huang and M. Bu\'enerd report ISN-01-18, March 2001;
               C.Y. Huang PhD thesis, University of Grenoble, Mai 2002.
\bibitem{HU03} C.Y. Huang, L. Derome, and M. Bu\'enerd, submitted to Astropart. Phys.
\bibitem{BYK}E Byckling and K. Kajantie, \textit{Particle kinematics},
                Wiley Interscience, 1973
\bibitem{TA76} F.E. Taylor \textit{et al}., {\ Phys. Rev. D.} {\bf 14}, 1217 (1976).
\bibitem{CO84} P.D.B.Collins and A.D.Martin, \textit{Hadron Interaction},
                Adam Hilger Ltd, Bristol, 1984
\bibitem{HO73}D. Horn and F. Zachariasen, \textit{Hadron Physics at Very High Energy},
                W.A. Benjamin, Inc. Advances Book Program, Massachusetts, 1973
\bibitem{LE83} J.R. Letaw \textit{et al}., {\ Astropart. J. Suppl. } {\bf 51}, 271 (1983).
\bibitem{MINUIT} F. James  {\ MINUIT } {\ Fonction Minimisation and Errore Analysis}, CERN,
                Program Library Long Writeup D506, 1998.
\bibitem{SU98} Y. Sugaya \textit{et al}., {\ Nucl. Phys. A }{\bf 634}, 115 (1998).
\bibitem{MA69} G.J. Marmer \textit{et al}., {\ Phys. Rev.} {\bf 179}, 1294 (1969).
\bibitem{AB93} T. Abott \textit{et al}., {\ Phys. Rev. D. } {\bf 47}, 1351 (1993).
\bibitem{AL70} J.V. Allaby \textit{et al}., {\ CERN-70-12} Nuclear Physics Division (1970).
\bibitem{DE65} D. Dekker \textit{et al}., {\ Phys. Rev.} {\bf 51}, 271 (1965).
\bibitem{EI72} T. Eichten \textit{et al}., {\ Nucl. Phys. } {\bf B44}, 333 (1972).
\bibitem{KA71} P.I.P. Kalmus \textit{et al}., {\ CERN-71-25} Nuclear Physics Division (1971).
\bibitem{SN85} J.M. Snow \textit{et al}., {\ Phys. Rev. D. }{\bf 32}, 11 (1985).
\bibitem{BA83} D.S. Barton \textit{et al}., {\ Phys. Rev. D. }{\bf 22}, 2580 (1983).
\bibitem{JO78} J.R. Johnson \textit{et al}., {\ Phys. Rev. D. } {\bf 17}, 1292 (1978).
\bibitem{BA74} W.F. Baker \textit{et al}., {\ Phys. Lett. } {\bf B51}, 303 (1974).
\bibitem{CR75} J.W. Cronin \textit{et al}., {\ Phys. Rev. D. } {\bf 11}, 3105 (1975).
\bibitem{AN79} D. Antreasyan \textit{et al}., {\ Phys. Rev. D. } {\bf 19}, 764 (1979).
\bibitem{GU76} K. Guettler \textit{et al}., {\ Nucl. Phys. } {\bf B116}, 77 (1976).
\bibitem{CA74} P. Capiluppi \textit{et al}., {\ Nucl. Phys. } {\bf B79}, 189 (1974).
\bibitem{AN71} Yu.M. Antipov \textit{et al}., {\ Phys. Lett. } {\bf B34}, 164 (1971).
\bibitem{AB84} V.V. Abramov \textit{et al}., {\ Z. phys. } {\bf C}, 205 (1984).
\bibitem{BO78} W. Bozzoli \textit{et al}., {\ Nucl. Phys. }{\bf B144}, 317 (1978).
\bibitem{BE98} I.G. Bearden \textit{et al}., {\ Phys. Rev. C. } {\bf 57}, 837 (1998).
\bibitem{TA82} L.C. Tan and L.K. Ng \textit{et al}, {\ Phys. Rev. D.}{\bf 26}, 1179 (1982).
\bibitem{DU03a} R.P. Duperray, K.V. Protasov, A.Yu. Voronin {\ Eur. Phys. J.}
                {\textbf A16}, 27 (2003)
\bibitem{DU03b} R.P. Duperray, K.V. Protasov, L. Derome and M.Bu\'enerd,
                submitted to {\ Eur. Phys. J.} /nucl-th 0301103
\bibitem{DU03c} R.P. Duperray et al., ICRC2003 Conf. Proc., Tsukuba (Japan), July 30 - Aug. 7, 2003.
\bibitem{AN73} M. Antinucci \textit{et al}., {\ Lettere Al Nuovo Cimento } {\bf 6}, 121 (1973).
\bibitem{GI01} G. Giacomelli \textit{et al}., {\ Il Nuovo Cimento } {\bf24C}, 575 (2001).

\end{thebibliography}
\end{document}